\documentclass[preprint,showpacs,preprintnumbers,amsmath,amssymb,prb]{revtex4}  
  
\usepackage{graphicx}  
\usepackage{dcolumn}  
\usepackage{bm}  
\usepackage{natbib}     
\usepackage{amssymb}     
\usepackage{epsfig}  
\usepackage{graphics}  
\usepackage{graphicx}  
     
\begin{document}     
  
\title{3d transition metal impurities in diamond:   
electronic properties and chemical trends}     
     
\author{L. V. C. Assali$^{\rm (1)}$, W. V. M. Machado$^{\rm (1)}$,  
and J. F. Justo$^{\rm (2)}$}  
  
\affiliation{$^{\rm (1)}$ Instituto de F\'{\i}sica,  
Universidade de S\~ao Paulo,\\  
CP 66318, CEP 05315-970, S\~ao Paulo, SP, Brazil \\  
$^{\rm (2)}$ Escola Polit\'ecnica, Universidade de S\~ao Paulo,\\  
CP 61548, CEP 05424-970, S\~ao Paulo, SP, Brazil}  
  
\begin{abstract}     
First principles calculations have been used to investigate the trends on 
the properties of isolated 3d transition metal impurities (from Sc to Cu) 
in diamond. Those impurities have small formation energies in the 
substitutional or double semi-vacancy sites, and large energies in the 
interstitial one. Going from Sc to Cu, the 3d-related energy levels in   
the bandgap move from the top of the bandgap toward the valence band in 
all three sites. Trends in electronic properties and transition energies
of the impurities,  in the substitutional or interstitial sites, 
are well described by a simple microscopic model considering the electronic 
occupation of the 3d-related levels. 
On the other hand, for the impurities in the double semi-vacancy site, 
there is a weak interaction between the divacancy- and the 3d-related 
orbitals, resulting in  in vacancy- and 3d-related levels in the materials 
bandgap.
\end{abstract}     
     
     
\maketitle  
         
\section{Introduction}     
\label{sec1}

Silicon-based device technology has flourished over the last four decades.
In such a time span, miniaturization was the keyword for improving device 
performance. When new challenges were foreseen in the horizon, designers 
used to find new solutions to overcome them \cite{end}. In a near 
future, intrinsic physical limits of this technology may preclude further 
improvements \cite{keyes}. A different route for electronic devices could 
be the use of wide bandgap semiconductors \cite{bun}, since, when compared to 
silicon, they present superior materials properties for electronic devices,
such as larger thermal conductivity, dielectric strength, and electron 
saturation velocity. There are still several technical limitations
that prevent their competitiveness with the well established silicon 
technology. However, those materials have found their niche over the 
last decade, with
applications in specific areas, such as high-power, high-temperature, 
high-frequency, opto-electronic \cite{gurbuz,wide}, and
spintronic \cite{dutt,kha} devices.

There is currently a high demand for devices to operate under extreme 
conditions, and diamond is one of the leading candidates for such 
applications \cite{hud,worth,nich}. This material has a wide bandgap 
(experimental value of 5.5 eV), high saturated carrier velocities, high 
electric field breakdown strength, low dielectric constant, and high thermal 
conductivity. Synthetic diamond has been grown out of graphite by high 
pressure-high temperature methods for about fifty years  \cite{bundy}  being 
currently the most widely used growing process that allows to get macroscopic 
diamond samples. In those processes, 3d transition metal (TM) alloys, 
involving mainly nickel, cobalt, and iron, are used as solvent-catalysts to 
both overcome the sp$^2$ to sp$^3$ energy barrier and accelerate the growth 
process \cite{yelisseyev}. Those TM end up being incorporated in the resulting 
diamond as residual impurities, either in isolated configurations or forming 
complexes with other defects, that can generate electrically and optically 
active centers \cite{yelisseyev,larico12}. Several transition metal-related 
active centers have been experimentally identified in diamond, and have been 
associated to impurities in substitutional, interstitial, 
or double semi-vacancy configurations \cite{larico12,collins1,iakou,nado}. 
Understanding the nature and microscopic structure of those centers is 
crucial in developing diamond-related technologies.

Here, we used first principles total energy calculations to investigate 
the electronic properties and chemical trends of 3d TM-related centers 
(from Sc to Cu) in diamond. We focused on the trends of impurities 
in the substitutional, interstitial, or double semi-vacancy sites. 
The stability of those defects, in all three sites, were computed in 
terms of their formation energies. The electronic structure of this 
3d$\rm ^n$ impurity family shows clear chemical trends in any site, with the 
3d-related levels deepening from the top of the gap toward the valence 
band with increasing number (n) of 3d electrons. Additionally, we show 
that the respective electronic properties and transition energies 
could only be rationalized in terms of the number of electrons 
occupying  the 3d-related energy levels within the materials bandgap.

\section{Methodology}     
\label{sec2}     
     
The calculations were carried using the all-electron spin-polarized 
full-potential linearized augmented plane wave (FP-LAPW) 
method \cite{singh}, implemented in the  WIEN2k package \cite{blaha}. 
The electron-electron interactions were described within the framework of 
the density functional theory and the generalized gradient 
approximation \cite{pbe}. Calculations were performed considering a 54-atom 
reference FCC supercell and a Monkhorst-Pack ($2 \times 2 \times 2$) 
grid to sample the irreducible Brillouin zone \cite{mp}. Convergence on 
the total energy was achieved using a plane wave basis set to describe the 
interstitial region, with the set limited by the wave number 7.0/R, where 
R = 1.2 a.u. (0.635 \AA) is the radius of all atomic spheres. 
Self-consistent iterations were performed until convergence on the total 
energy of 10$^{-4}$ Ry was achieved. In all systems with
impurities, the internal degrees of freedom 
were optimized, without any symmetry constraints, until the force in 
each atom was smaller than 10$^{-3}$ Ry/a.u. Such theoretical framework and convergence 
criteria have been shown to provide a reliable description of the 
electronic properties of defect centers in several 
semiconductors \cite{assali,larico}.

The formation energy of a TM impurity center in diamond (${E}_{f}^q$) 
was computed by \cite{assali,larico12}:  
\begin{equation}  
{E}_{f}^q =  E^q_{tot}(N_{\rm C},N_{TM}) - N_{\rm C} \mu_{\rm C} -
N_{TM}\mu_{TM}+q(\varepsilon'_v + \epsilon_F) \, ,    
\label{eq1}  
\end{equation}  
where $E^q_{tot}(N_{\rm C},N_{TM})$ is the total energy of a supercell 
with the defect in a $q$ charge state, with $N_{\rm C}$ carbon atoms 
and $N_{TM}$ TM impurity atoms. The $\mu_{\rm C}$ and $\mu_{TM}$ are the chemical 
potentials of respectively diamond and TM stable crystalline structures,
computed within the same methodology described in the previous paragraph.  
Additionally, $\epsilon_F$ is the Fermi energy  
($0 \leq \epsilon_F \leq \varepsilon_g$, where $\varepsilon_g$ is the bandgap energy ) 
and $\varepsilon'_{v}= \varepsilon_{v} + \delta_{q}$, where
$\delta_{\rm q}$ is a parameter that lines up the band structures of the 
bulk diamond material with and without the impurity in relation to the 
top of valence band of the pure crystal ($\varepsilon_{v}$). 
Within our theoretical approximations,
we found $\varepsilon_g = 4.56$ eV for diamond.

For a certain impurity (in a substitutional, interstitial, or double 
semi-vacancy center), the transition energy between the $q^\prime$ and $q$ 
charge states, $E_t (q^\prime /q)$, is the value of the Fermi energy in the bandgap 
when the formation energies of the center in those respective
charge states, as computed by equation \ref{eq1}, are equal. 
\begin{equation}  
{E}_{t} (q^\prime /q) = \epsilon_F \ \ \ {\rm when} \ \ \  
{E}_{f}^{q^\prime} -  {E}_{f}^q = 0 \, . 
\label{eq2}
\end{equation}
We should stress that here we computed the transition energies, not the
excitation ones, in which other methodologies could be more 
appropriate \cite{albu}.

\section{Results and Discussion}     
\label{sec4}     
       
We computed the properties of the 3d transition metal impurities in diamond,
from Sc (atomic configuration $\rm 3d^1 4s^2$) to Cu (atomic configuration 
$\rm 3d^{10} 4s^1$), in three different sites: tetrahedral interstitial 
(TM$_{\rm i}$), substitutional (TM$_{\rm s}$), and double semi-vacancy 
(TM$_{2V}$) ones. 

\begin{figure}[h]  
\centering{  
\includegraphics[width=150mm]{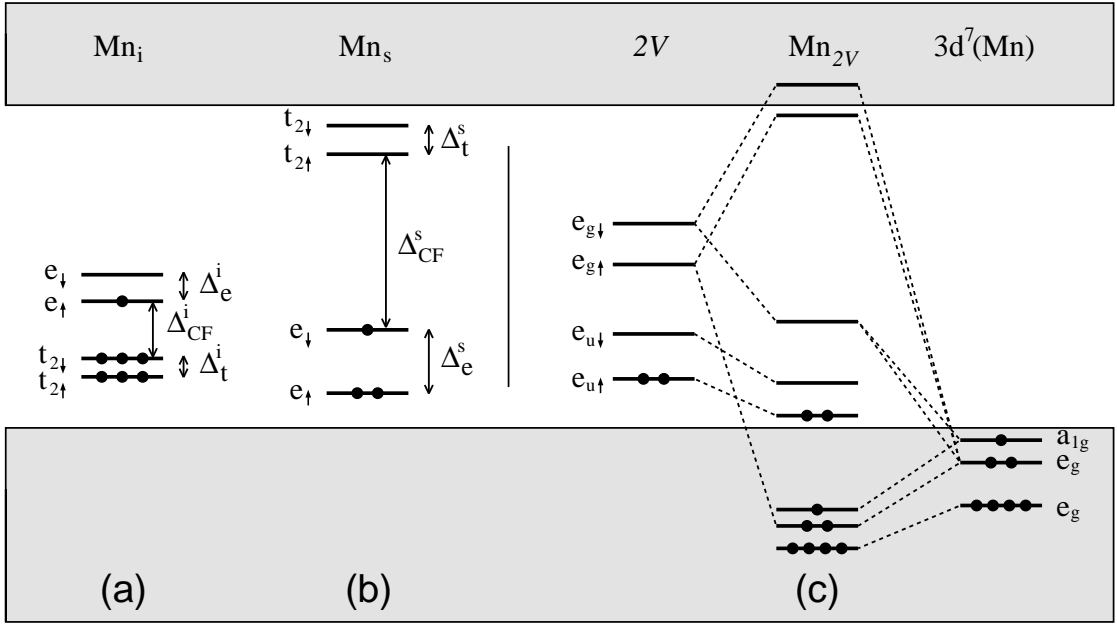}  
\caption{Schematic representation of the gap energy electronic states for
a manganese impurity in (a) interstitial (Mn$_{\rm i}$), (b) substitutional 
(Mn$_{\rm s}$) and (c) double semi-vacancy (Mn$_{2V}$) sites.
In the case of (Mn$_{2V}$), the figure presents the model of a hybridization
between the divacancy states ($2V$) and the 3d electrons
in an isolated atomic configuration (Mn).
The $\uparrow$ and $\downarrow$ arrows represent the spin up and
down, respectively. Gray regions represent the diamond valence and conduction
bands. For simplicity, the systems are represented considering
a tetrahedral symmetry, neglecting symmetry lowering distortions.
For Mn$_{i}$ (Mn$_{s}$), the figure presents the 
exchange potential splittings
$\Delta_{\rm e}^{\rm i}$ and $\Delta_{\rm t}^{\rm i}$ 
($\Delta_{\rm e}^{\rm s}$ and $\Delta_{\rm t}^{\rm s}$) 
and the crystal field splitting
$\Delta_{\rm CF}^{\rm i}$ ($\Delta_{\rm CF}^{\rm s}$).}
\label{fig1}  
}  
\end{figure}

When a TM impurity, with a 3d$^{\rm n}$4s$^2$ ($\rm 1\leq n < 9$) atomic 
configuration, occupies a tetrahedral interstitial site in diamond (TM$_{\rm i}$), 
its 4s electrons are transferred to the 3d orbitals, resulting in a 
3d$^{\rm n+2}$ configuration. In a tetrahedral crystal field, the 3d 
states are split into $e+t_2$ irreducible representations, with the 
threefold $t_2$ states lying lower in energy than the two-fold $e$ ones. 
This level ordering is the result of a strong octahedral crystal field created 
by the next nearest neighboring carbon atoms, that leads to a strong
repulsion in the $e$ states, moving them up, in the upper half of the bandgap.  
Additionally, 
the crystal field energy splitting ($\Delta_{\rm CF}^{\rm i}$) is larger 
than the exchange splittings ($\Delta_{\rm e}^{\rm i}$ and 
$\Delta_{\rm t}^{\rm i}$), such that the system always presents a low spin 
configuration. This results from the small lattice parameter of diamond, and
is consistent with TM impurities in boron nitride \cite{bn}.
Figure \ref{fig1}(a) presents the specific case of 
Mn$_{\rm i}$ in the neutral charge state (3d$^7$). For other TM$_{\rm i}$ 
impurities in diamond (from Sc to Cu), their $t_2$ and $e$ levels have  
an equivalent behavior to 
those of Mn$_{\rm i}$, but with a chemical trend such that those levels move 
all together from the midgap region, in Sc$_{\rm i}$, toward the valence 
band, in Cu$_{\rm i}$, consistent with results for TM$_{\rm i}$ impurities in 
other semiconductors \cite{assali0,assali09}. 
The 3d-character localization 
in the TM atomic spheres increases with increasing the atomic number.

In the substitutional site (TM$_{\rm s}$), the impurity presents a
3d$^{\rm n-2}$ configuration, since four electrons are necessary for a 
covalent bonding with the four nearest neighboring carbon atoms. In contrast
to the interstitial case, the tetrahedral crystal field  drives the $e$ 
states to lie lower in energy than the $t_2$ ones. Here, the crystal field splitting 
also prevails over the exchange splittings  
($\rm \Delta_{CF}^{s}\gg  \Delta_{t}^{s}\ {\rm and} \ \Delta_{e}^{s}$). 
Figure \ref{fig1}(b) presents the 
specific case of Mn$_{\rm  s}$ in the neutral charge state (3d$^3$). For 
other TM$_{\rm s}$ impurities in diamond (from Sc to Cu), the $e$ and $t_2$ 
levels are equivalent to those of Mn$_{\rm s}$, but with a chemical trend 
such that those levels move all together from the bandgap top, in 
Sc$_{\rm s}$, toward the valence band, in Cu$_{\rm s}$ \cite{assali09}.
The 3d-character localization 
in the TM atomic spheres also increases with increasing the atomic number.

The electronic structure of the TM impurity in the double semi-vacancy site
(TM$_{2V}$) is considerably more complex than that in the other two sites.
The electronic structure of TM$_{2V}$ centers are result of an interaction 
between the divacancy states with the ones coming from the atomic TM, 
as represented in figure \ref{fig1}(c) for the Mn$_{2V}$ center. 
The one-electron ground state structure of a diamond 
divacancy in D$_{3d}$ symmetry has the e$_u^2$e$_{g}^0$ configuration 
in the bandgap region. In that symmetry, the TM 3d-related energy 
levels are split into 2e$_g$+a$_{1g}$. When a TM atom is placed in the 
middle position of a divacancy, one of its e$_g$ energy levels interacts slightly 
with the carbon dangling bonds, leaving a fully occupied non-bonding 
t$_{2g}$-like (e$_g$ + a$_{1g}$) orbital inside the valence band. However, 
the other TM-related e$_g$ orbital interacts with the divacancy-related e$_g$ gap 
state, resulting in an e$_g$-bonding level in the valence band and  
an e$_g$-anti-bonding one in the bandgap. The divacancy-related e$_{\rm u}$ 
orbitals do not interact with any 3d-related TM ones, and consequently 
they remain near the top of the valence band.

\begin{figure}[ht]  
\centering{  
\includegraphics[width=137mm]{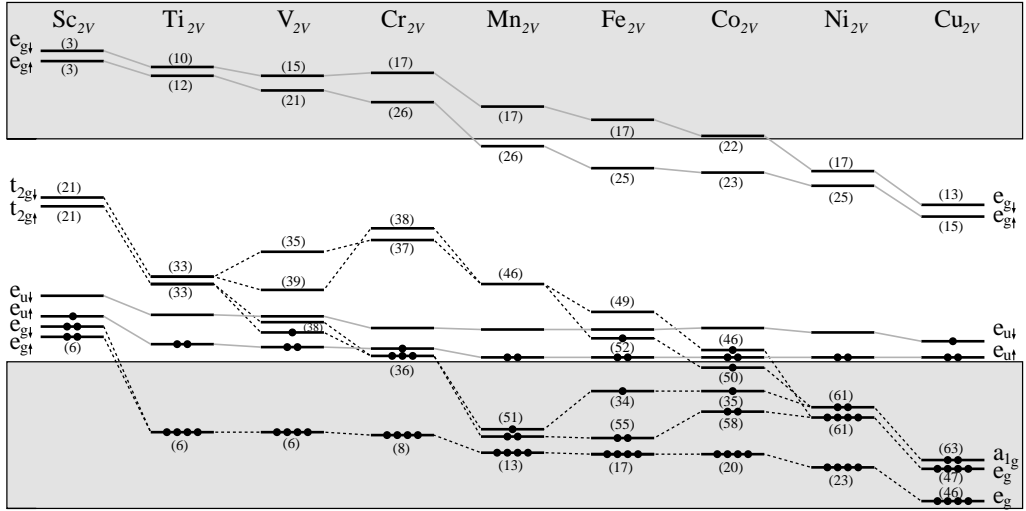}  
\caption{The energy eigenvalues representing the divacancy- and TM 
3$d$-related levels in the gap region for the TM$\rm _{2V}$ centers 
(TM = Sc, Ti, V, Cr, Mn, Fe, Co, Ni, and Cu). Levels with spin up and 
down are represented by $\uparrow$ and $\downarrow$ arrows, respectively.   
The filled  (open) circles represent the electronic (hole)   
occupation of the gap levels. Numbers in parenthesis represent the  
$d$-character percentage of charge inside the TM atomic spheres.
Level labeling is consistent with the model presented in Fig. \ref{fig1}(c).
For clarity, the results presented in the figure corresponded to a
high-symmetry D$_{3d}$ configuration, although the converged calculations took
no symmetry constraints, as presented in Table \ref{tab1}.
Levels associated to the TM (divacancy) are connected by dashed (full gray) 
lines.}  
\label{fig2}  
}  
\end{figure}

For the TM$_{2V}$ center, depending on the impurity atomic number, the 
relative position of the 3d-related levels and
the divacancy related ones may switch along the series, such that the 
electronic structure of each impurity depends strongly on such relative positions
and the center charge state.
Figure \ref{fig2} presents the electronic structure of the TM$_{2V}$, in the 
neutral charge state. The figure presents, in parenthesis, the 3d character 
inside the TM atomic sphere of each energy level. Going from Sc$_{2V}$ to
Cu$_{2V}$ impurity centers, the 3d-t$_{\rm 2g}$-related 
levels move from the middle of the bandgap toward the valence band, crossing 
with the divacancy-related e$_{\rm u}$ ones, that lie near the valence band top.  
Additionally, the percentage of d-character increases along the series. 
These trends are consistent with those for substitutional and interstitial TM impurities.
However, the electronic character of the highest occupied level of those centers
 depends on the relative position of the
divacancy-related levels with respect to the 3d-related ones.
According to figure \ref{fig2} for the TM$_{2V}$ in the neutral charge state, 
it is associated to the TM for V, Fe, and Co, while it 
is associated to the divacancy for Sc, Ti, Cr, Mn, Ni, and Cu.

\begin{figure}[h!] 
\centering{  
\includegraphics[width=100mm]{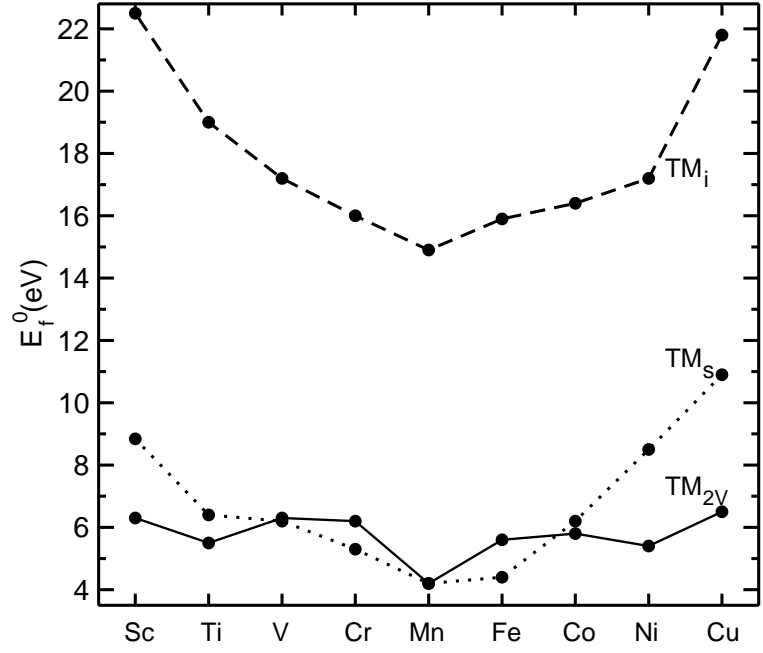}  
\vspace*{0.7cm}
\caption{Formation energy of the 3d TM impurities,
in the neutral charge state (E$_{\rm f}^0$), in   
substitutional (dotted line), interstitial (dashed line), and  
double semi-divacancy (full line) sites.}   
\label{fig3}  
}  
\end{figure}

Figure \ref{fig3} presents the trends on the formation energy for neutral 
impurities in all three sites, as computed by equation \ref{eq1}. 
The results show that the formation energies of a TM with 3d$^{\rm n}$ or 
3d$^{\rm 10-n}$ configurations are essentially equivalent in any site, with 
a clear energy favoring for the impurity in the middle of that family, which 
is manganese. Additionally, the interstitial site is the most unfavorable one 
for any TM, with formation energies of  more than 10 eV higher than the 
respective ones in the other two  sites \cite{larico}. The small energy 
difference for the
impurity in the substitutional or double semi-vacancy sites suggests that 
those two configurations may compete, co-existing in the diamond samples.  
We should also stress that trends in energy differences between 
TM defects in interstitial and substitutional configurations remain
essentially the same for defects in charge states other than neutral.
Those trends in energy are consistent, for example, with available 
experimental data for concentrations of cobalt \cite{yelisseyev} and
nickel \cite{nado,larico12} impurities in as-grown and annealed 
synthetic diamond.

Table \ref{tab1} presents the local symmetry and spin of the relaxed 
configuration for the TM impurities in the neutral charge state. 
As result of the electronic structures, the TM$_{2V}$ centers generally 
present spin values that are larger than the TM$_{\rm i}$ and TM$_{\rm s}$ 
centers. The manganese in the double semi-vacancy presents the 
largest spin of all centers (S=5/2). This large spin value suggests potential 
applications for spintronic devices \cite{jung,lombardi}.

Up to now, we have discussed the properties of the TM impurities in their
neutral charge state. However, since those impurities introduce
energy levels in the bandgap, both occupied and unoccupied ones, there is 
often a large number of possible stable charge states for each center.  
The transition energy between two different 
charge states is an important information for experimentalists to identify a
certain center. Here, we carried out calculations for all stable charge states 
of each TM impurity in those three sites. Some centers presented up to 
seven different stable charge states, with transition energies lying in the 
diamond bandgap. The transition energy between two different charge states 
of a certain center was computed using the total energies of 
the centers in the initial and final
charge states, as  given by equations  \ref{eq1} and \ref{eq2}.

\begin{figure}[h!]  
\centering{  
\includegraphics[width=150mm]{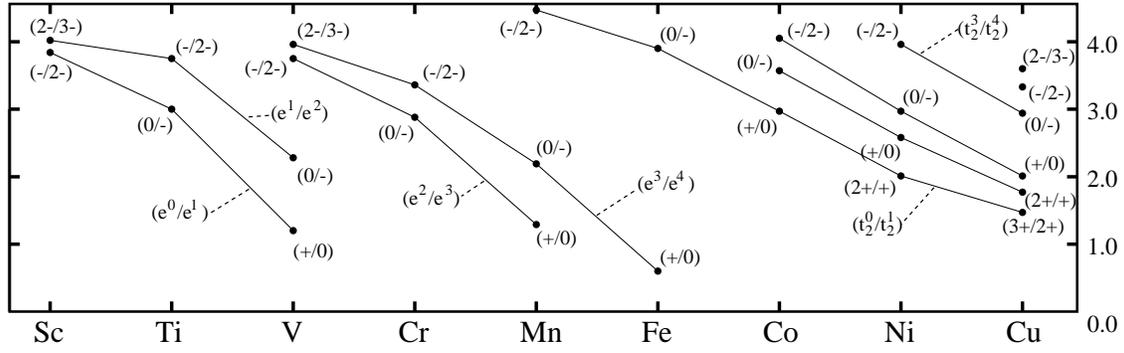}
\caption{Transition energies (E$_{\rm t}$) of 3d TM impurities
in the substitutional site. The initial and final 3d-related 
bandgap electronic configurations are given in square brackets
(see text). The lines in the figure are only guides
to the eye, connecting different transition states related with the 
same electronic configurations of different 
TM impurities.}
\label{fig4}  
}  
\end{figure}

Our results indicated that the chemical trends on transition 
energies (computed by equation \ref{eq2})
along the 3d series,  could only be rationalized if they were discussed in 
terms of the 3d-related level occupation and the respective crystal field 
and exchange splittings, as presented in figure \ref{fig1}. 
The comparison among different TM impurities should be 
conducted considering the same number of total electrons in each system, as 
discussed below.  The transition energies within the diamond bandgap for 
TM$_{\rm s}$ (from Sc to Cu) are shown in figure \ref{fig4}. 
The stable charge state of a certain center depends on the
position of the Fermi level in the bandgap. For example, the
Ti$_{\rm s}$ presents two transition states in the bandgap, the (0/-) and
(-/2-), as shown in figure \ref{fig4}. This means that this center  
is stable in the neutral charge state for $0<\epsilon_F \leq 3.0$~eV,
in the negatively charge state for $3.0 \leq \epsilon_F \leq 3.8$~eV, and 
in the double negatively charge state for $3.8\,\, {\rm eV} \leq \epsilon_F < \varepsilon_g$.

According to figure \ref{fig1}(b), the TM$_{\rm s}$ impurity centers 
introduce energy levels with $e$ and $t_2$ irreducible representations,
with $e_\uparrow$ and $e_\downarrow$ states below $t_{2\uparrow}$ and
$t_{2\downarrow}$ ones. For an impurity with a 3d$^{\rm n}$4s$^2$ atomic
configuration, those levels are filled with $\rm (n-2)$d-electrons. 
For Sc$_{\rm s}^{-}$ center ($\rm n=2$),
the $e$ and $t_2$ states
are empty and the bandgap electronic configuration  is
$e^0_\uparrow e^0_\downarrow t_{2\uparrow}^0 t_{2\downarrow}^0$,
where the level ordering is consistent with the increasing values of the respective
energy eigenvalues, as given in figure \ref{fig1}(b). 
The Sc$_{\rm s}^{2-}$ center is related to adding one electron 
to an unoccupied $e$ level, leading to the 
$e^1_\uparrow e^0_\downarrow t_{2\uparrow}^0 t_{2\downarrow}^0$
bandgap electronic configuration. 
Therefore, the $(-/2-)$ transition state is  
described by the connection  between initial and final 
electronic configurations, as represented by 
$e^0_\uparrow e^0_\downarrow t_{2\uparrow}^0 t_{2\downarrow}^0  \rightarrow 
e^1_\uparrow e^0_\downarrow t_{2\uparrow}^0 t_{2\downarrow}^0 $. It could be
represented in a compact form  as $e^0 \rightarrow e^1$, or  
just [$e^0 / e^1$] in figure \ref{fig4}.  
Accordingly, the $(2-/3-)$ transition state, or [$e^1/ e^2$], 
is related to the  $e^1 \rightarrow e^2$ electronic configurations. 
For Sc$_{\rm s}$, 
there is no additional transition state that lies in the bandgap.

For Ti$_{\rm s}$, the ($0/-$) and ($-/2-$) transition states 
 are associated 
with  $e^0_\uparrow  \rightarrow e^1_\uparrow$ 
and $e^1_\uparrow  \rightarrow e^2_\uparrow$  electronic configurations, 
respectively. Those two transition states should be compared 
respectively to the  
$(-/2-)$ e $(2-/3-)$ transition ones of Sc$_{\rm s}$ center, that is why
the transitions are connected by lines, as a guide to the eye, in figure \ref{fig4}..
Going to V$_{\rm s}$, the same electronic configurations are observed, 
being now related to the  ($+/0$) and ($0/-$) transition energies.  
However, the V$_{\rm s}$ center carried more two possible electronic
configurations in the diamond bandgap, related to   V$_{\rm s}^{2-}$ 
and  V$_{\rm s}^{3-}$ centers, allowing to compute the transition energies 
between the  electronic configurations  
$e^2_\uparrow e^0_\downarrow t_{2\uparrow}^0 t_{2\downarrow}^0 
\rightarrow e^2_\uparrow e^1_\downarrow t_{2\uparrow}^0 t_{2\downarrow}^0 $ 
(or in a compact form, $e^2 \rightarrow e^3$) and $e^3 \rightarrow e^4$, 
associated with the ($-/2-$) and  ($2-/3-$) transition states, respectively. The 
Mn$_{\rm s}$ impurity has four stable charge states in the bandgap. 
For this impurity, after the $e_\uparrow$ and  $e_\downarrow$ states 
are fully occupied, additional electrons could only occupy the $t_2$ states. 
For Mn$_{\rm s}$, there is the  
$e^4t_{2\uparrow}^0t_{2\downarrow}^0  \rightarrow
e^4t_{2\uparrow}^1t_{2\downarrow}^0$  transition, associated with the ($-/2-$) one. 
From Co$_{\rm s}$ to Cu$_{\rm s}$, there are
$ e^4t_{2\uparrow}^0 t_{2\downarrow}^0 \rightarrow
e^4t_{2\uparrow}^1t_{2\downarrow}^0 $,  $ e^4t_{2\uparrow}^1t_{2\downarrow}^0
\rightarrow  e^4t_{2\uparrow}^2t_{2\downarrow}^0 $,
$ e^4t_{2\uparrow}^2 t_{2\downarrow}^0 \rightarrow
e^4t_{2\uparrow}^3t_{2\downarrow}^0 $, and  
$ e^4t_{2\uparrow}^3 t_{2\downarrow}^0 \rightarrow  
e^4t_{2\uparrow}^3 t_{2\downarrow}^1 $
(or $t_{2}^3 \rightarrow t_{2}^4$) transitions. Therefore, one could
associate, for example, the ($-/2-$) transition of Sc$_{\rm s}$ with the
($0/-$) one of Ti$_{\rm s}$ and the ($+/0$) of V$_{\rm s}$, i.e. transitions
that involve the same number of
electrons in the system. For all those centers, it is observed that any
transition state has a chemical trend to move from the top of the bandgap 
in Sc toward the valence band maximum in Cu. Such trend is result of a Coulomb 
interaction, since the increasing atomic number, with a constant number of
electrons, increases the nuclei-electron 
attractive interaction, reducing the total energy of the system.

Those results allow to establish  a microscopic model for a transition energy, 
associated with the 3d level occupation in the bandgap. There is a small
energy difference between the $e^0 \rightarrow e^1$ and  $e^1 \rightarrow e^2$ 
transitions, which is associated with the occupation of the $e_\uparrow$, 
being result of a Coulomb interaction. Now, there is a large energy difference 
between  $e^1 \rightarrow e^2$  and $e^2 \rightarrow e^3$ transitions, which 
results from the exchange potential splitting ($\Delta_{\rm e}^{\rm s}$). 
This is because the $e^2 \rightarrow e^3$ transition differs from  the 
$e^1 \rightarrow e^2$ one by the addition of an electron with a different spin.  
The large energy difference between  the $e^3 \rightarrow e^4$ and 
$t_2^0 \rightarrow t_2^1$ transitions results from the presence of
a strong crystal field 
splitting ($\Delta_{\rm CF}^{\rm s}$), which is larger than the exchange 
potential splittings ($\Delta_{\rm e}^{\rm s}$ and $\Delta_{\rm t}^{\rm s}$).
For the transitions related to the occupation of the $t_{2\uparrow}$ state
(up to 3 electrons), the small energy difference is result of a Coulomb interaction.
Finally, for the  $t_2^3 \rightarrow t_2^4$ transition, the
large energy difference is controlled by the exchange potential splitting of 
the $t_2$ state ($\Delta_{\rm t}$).

\begin{figure}[h!]  
\centering{  
\includegraphics[width=150mm]{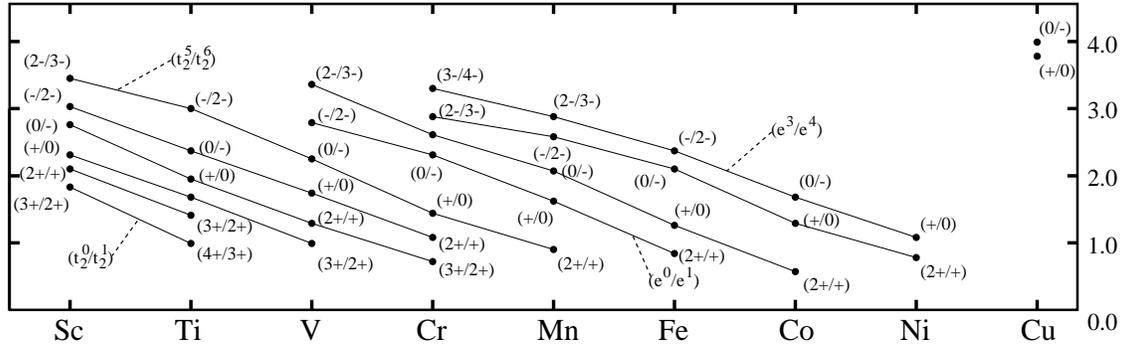}
\caption{Transition energies (E$_{\rm t}$) of 3d TM impurities
in the interstitial site. The lines and symbols 
are consistent with the ones in figure \ref{fig4}.}   
\label{fig5}  
}  
\end{figure}

The same model could be used to discuss the transition energies associated with 
TM impurities in the interstitial site, as presented in figure \ref{fig5}. 
According to figure \ref{fig1}(a),  
the TM$_{\rm i}$ centers introduce energy levels with $t_2$ and $e$  irreducible 
representations in the bandgap, with $t_{2\uparrow}$ and $t_{2\downarrow}$ 
states below the $e_\uparrow$ and $e_\downarrow$ ones. For an impurity with a 
3d$^{\rm n}$4s$^2$ atomic configuration, those levels are filled with 
$(n+2)$d-electrons. The first possible transition energy  is associated to 
the electronic initial and final configurations  
$t_{2\uparrow}^0 t_{2\downarrow}^0 e_{\uparrow}^0 e_{\downarrow}^0 
\rightarrow t_{2\uparrow}^1 t_{2\downarrow}^0 e_{\uparrow}^0 e_{\downarrow}^0$.
It could be represented in a compact form as $t_2^0 \rightarrow t_2^1$, or just 
[$t_2^0 / t_2^1$] in figure \ref{fig5}. Those configurations are associated
with the (3+/2+) and (4+/3+) transition states for the Sc$_{\rm i}$ 
and Ti$_{\rm i}$ impurities, respectively. 
After all the $t_2$ states are filled with increasing Fermi energy, 
the large energy difference for the $t_{2}^6e^0_\uparrow e^0_\downarrow  \rightarrow
t_{2}^6 e^1_\uparrow e^0_\downarrow$ electronic configurations 
(or in compact form $e^0  \rightarrow e^1$) is associated with
the crystal field splitting ($\Delta_{\rm CF}^{\rm i}$), consistent with 
the model of figure \ref{fig1}.
For all those centers, it is observed that any
transition state, along the 3d series,  has a chemical trend
to move from the upper half region of the bandgap, in Sc,  toward the 
valence band maximum, in Cu. Additionally, the crystal field splitting 
for the TM$_{\rm i}$ centers is considerably smaller than the respective 
splitting for the TM$_{\rm s}$ ones ($\Delta_{\rm CF}^{\rm i} \ll 
\Delta_{\rm CF}^{\rm s}$).

\begin{figure}[h!]  
\centering{  
\includegraphics[width=150mm]{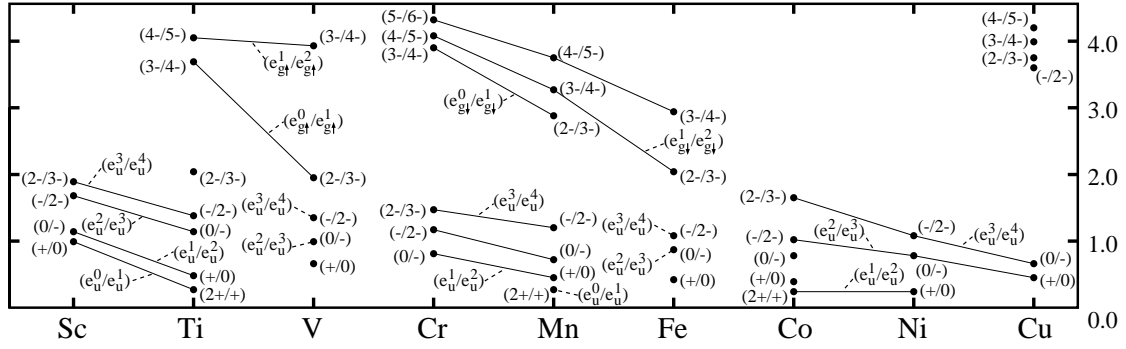}
\caption{Transition energies (E$_{\rm t}$) of 3d TM impurities
in the double semi-vacancy site.The lines and  symbols
are consistent with the ones in figure \ref{fig4}.}   
\label{fig6}  
}  
\end{figure}

For the TM$_{2V}$, this model could be applied only to a few transition states,
since the almost non-interacting divacancy-related energy levels and the 
3d-related ones lie in the same region of the bandgap, 
as shown in figure \ref{fig2}. 
As mentioned earlier, the divacancy-related states ($e_{\rm u}$)
remain in the bottom of the bandgap, the 3d-related states 
($e_{\rm g}+a_{\rm  1g}$) move from the top of the gap in Sc toward the
valence band in Cu. While some transition states are computed in 
association with electronic configurations 
filling the divacancy-related states, others are
associated with filling the TM-related states. The transition energies 
associated with the  TM$_{2V}$ centers are presented in figure \ref{fig6}.
The trends can be here observed for some TM impurities. On the other hand,
the V, Fe, and Co are exceptions, since the highest occupied level in
their neutral charge state has a 3d-related character, as shown in figure
\ref{fig2}. For those cases, the electronic configurations are considerably
richer, which are presented in table \ref{tab2} for several charge states. 
Therefore, table \ref{tab2} helps to assign the electronic configurations 
associated with the respecive transitions.
According to figure \ref{fig6}, in general, transition energies related to 
occupying the divacancy-related orbitals ($e_{\rm u}$) are in the lower part 
of the bandgap while transition states associated to occupying the 
3d-related states ($e_{\rm g}$ and
$a_{\rm 1g}$)  are mostly in the upper part of the bandgap.

\section{Summary}     
\label{sec5}     
       
In summary, we have investigated the electronic properties and
chemical trends of isolated 3d-transition metal impurities in 
diamond. We have shown that impurities in the 
substitutional or double semi-vacancy centers have smaller
formation energies than the isolated centers in the interstitial site.
We have also shown that trends on transition energies,
for any of the three sites, could only be rationalized if they
were discussed in terms of the increasing occupation of 3d-related
states in the bandgap. Such trends are consistent with what would
be expected for transition metal impurities in other semiconductors
in either isolated configurations \cite{rae,conti} 
or forming complexes with other defects \cite{assali0,zhao}.

\vspace*{0.8cm}

\noindent  
{\bf Acknowledgments:}     
The authors acknowledge support from Brazilian agencies CNPq and FAPESP. The     
calculations were performed in part using the computational facilities of 
the CENAPAD and the LCCA-CCE of the University of S\~ao Paulo.

\pagebreak

\begin{table}[ht]  
\caption{Point symmetry and spin (S) 3d TM impurities   
in the neutral charge state.  The table presents results for interstitial, 
substitutional and double semi-vacancy sites.}  
\label{tab1}  
\vspace*{0.5cm}  
\begin{center}  
\begin{tabular}{ccccccccccc}  
\hline \hline  
\ \ \ Site \ \ \ & \ \ \ \ \ \ \  & \ \ Sc \ \ & \ \ Ti \ \ & \ \ V \ \ & 
\ \ Cr \ \ & \ \ Mn \ \ & \ \ Fe \ \ & \ \ Co \ \ & \ \ Ni \ \ & \ \ Cu\ \ \\  
\hline  
TM$_{\rm i}$  & Sym. & T$_{\rm  d}$ & D$_{\rm 2d}$ & D$_{\rm 2}$ & T$_{\rm  d}$ &   
D$_{\rm 2d}$  &  T$_{\rm  d}$ & D$_{\rm 2d}$ & T$_{\rm  d}$ &  D$_{\rm 2d}$ \\  
             & S     & 3/2 & 0 & 1/2 & 0 & 1/2 & 1 & 1/2 & 0 & 1/2 \\  

TM$_{\rm s}$  & Sym. & D$_{\rm 2d}$  & T$_{\rm  d}$ & D$_{\rm 2d}$ & T$_{\rm  d}$ &   
D$_{\rm 2d}$  &  T$_{\rm  d}$ & D$_{\rm 2d}$ & C$_{\rm 1}$  & T$_{\rm  d}$  \\  
             & S & 1/2 & 0 & 1/2 & 1 & 1/2 & 0 & 1/2 & 1 & 3/2 \\   
TM$_{\rm 2V}$ & Sym. & C$_{\rm 2h}$  & D$_{\rm 3d}$ & D$_{\rm 3d}$ & C$_{\rm 2h}$   
& D$_{\rm 3d}$  & D$_{\rm 3d}$  & C$_{\rm 2h}$ &  D$_{\rm 3d}$ & C$_{\rm i}$\\  
             & S  & 3/2 & 1 & 3/2 & 2 & 5/2 & 2 & 3/2 & 1 & 1/2 \\  
\hline \hline  
\end{tabular}  
\end{center}  
\end{table}  
\pagebreak

\begin{table}[h]
\caption{Electronic configuration (EC), for the divancancy-related  and
TM 3d-related energy levels, and spin (S) for the TM$_{2V}^q$ centers
in diamond in $q$ charge state, with $\rm TM = V, \,Fe,~ \mbox{and}~ Co$.}
\label{tab2}  
\vspace*{0.5cm}  
\begin{center}
\begin{tabular}{lcccccc}
\hline \hline   
\multicolumn{1}{l}~~{$q$}~~ & \multicolumn{2}{c} {~~~~~~~~V$_{2V}$} &
\multicolumn{2}{c} {~~~~~~~~Fe$_{2V}$} & \multicolumn{2}{c} {~~~~~~~~Co$_{2V}$}  \\
  &EC & ~~~S~~~ & EC & ~~~S~~~ & EC & ~~~S~~~   \\ \hline
 $2+$ & ------ & --- & ------ & --- & $e_{g \uparrow}^2 a_{1g\uparrow}^1
 a_{1g\downarrow}^1 e_{u \uparrow}^1 e_{g \downarrow}^0  e_{u\downarrow}^0 $ & 3/2\\
 $+$ & $e_{u\uparrow}^2  a_{1g\uparrow}^0 e_{g\uparrow}^0
a_{1g\downarrow}^0  e_{u\downarrow}^0 e_{g\downarrow}^0$ & 1 &
$a_{1g\uparrow}^1  e_{g\uparrow}^2  e_{u \uparrow}^2   a_{1g \downarrow}^0e_{g\downarrow}^0 e_{u\downarrow}^0 $ &
5/2 &

$e_{g \uparrow}^2 a_{1g\uparrow}^1
 a_{1g\downarrow}^1 e_{u \uparrow}^2e_{g \downarrow}^0  e_{u\downarrow}^0$
& 2 \\
$0$  &$e_{u\uparrow}^2 a_{1g\uparrow}^1  e_{g\uparrow}^0 e_{u\downarrow}^0
 e_{g\downarrow}^0 a_{1g\downarrow}^0 $ & 3/2 &
 $  e_{g\uparrow}^2 a_{1g\uparrow}^1 e_{u \uparrow}^2  a_{1g\downarrow}^1 e_{u
   \downarrow}^0 e_{g\downarrow}^0$ &
2  & $e_{g \uparrow}^2 a_{1g\uparrow}^1
 a_{1g\downarrow}^1 e_{u \uparrow}^2e_{g \downarrow}^1  e_{u\downarrow}^0$ & 3/2 \\
 $-$  &$e_{u\uparrow}^2  a_{1g\uparrow}^1 e_{u\downarrow}^1 e_{g\uparrow}^0
e_{g\downarrow}^0 a_{1g\downarrow}^0 $ & 1 &
$e_{g\uparrow}^2  a_{1g\uparrow}^1 e_{u \uparrow}^2  a_{1g\downarrow}^1 e_{u
   \downarrow}^1 e_{g\downarrow}^0$ & 3/2 &
 $ e_{u \uparrow}^2  a_{1g\uparrow}^1
 a_{1g\downarrow}^1 e_{g \uparrow}^2 e_{g \downarrow}^2  e_{u\downarrow}^0$
&   1 \\
 $2-$  &$e_{u\uparrow}^2 e_{u \downarrow}^2  a_{1g\uparrow}^1  e_{g\uparrow}^0
e_{g\downarrow}^0 a_{1g \downarrow}^0$ & 1/2 &
$e_{g\uparrow}^2  a_{1g\uparrow}^1 e_{u \uparrow}^2 e_{u
   \downarrow}^2 a_{1g\downarrow}^1  e_{g\downarrow}^0$ & 1 &
$e_{u\uparrow}^2 a_{1g\uparrow}^1 a_{1g\downarrow}^1 e_{g\uparrow}^2
e_{g\downarrow}^2e_{u\downarrow}^1$  & 1/2  \\
 $3-$  &$e_{u\uparrow}^2 e_{u \downarrow}^2  e_{g\uparrow}^2 a_{1g\uparrow}^0
a_{1g \downarrow}^0e_{g\downarrow}^0 $ &  1 &
$e_{g\uparrow}^2 e_{u \uparrow}^2  e_{u
   \downarrow}^2 a_{1g\uparrow}^1   e_{g\downarrow}^2 a_{1g\downarrow}^0$ & 1/2 &
$e_{u \uparrow}^2e_{u \downarrow}^2  a_{1g\uparrow}^1 a_{1g
   \downarrow}^1 e_{g \uparrow}^2  e_{g \downarrow}^2  $ &  0 \\
 $4-$  & $e_{u\uparrow}^2 e_{u \downarrow}^2  e_{g\uparrow}^2 a_{1g\uparrow}^1
e_{g\downarrow}^0 a_{1g \downarrow}^0$  & 3/2
 & $e_{u \uparrow}^2e_{u \downarrow}^2  a_{1g\uparrow}^1 a_{1g
   \downarrow}^1 e_{g \uparrow}^2  e_{g \downarrow}^2  $  & 0 & ------   & --- \\
\hline \hline  
\end{tabular}
\end{center}
\end{table}
\pagebreak

 \pagebreak

\pagebreak

 \pagebreak

\pagebreak

 \pagebreak

\end{document}